\documentclass[aps,prl,twocolumn,superscriptaddress,floatfix,amsmath,showpacs]{revtex4}

\usepackage{graphicx}

\begin{document}
\title{Reverse Doppler effect in backward spin waves scattered on acoustic waves}

\author{A. V. Chumak}
\affiliation{Fachbereich Physik and Forschungszentrum OPTIMAS,
Technische Universit\"at Kaiserslautern, 67663 Kaiserslautern,
Germany}
\author{P. Dhagat}
\affiliation{School of Electrical Engineering and Computer Science,
Oregon State University, Corvallis, OR, USA}
\author{A. Jander}
\affiliation{School of Electrical Engineering and Computer Science, Oregon State University, Corvallis,
OR, USA}
\author{A. A. Serga}
\affiliation{Fachbereich Physik and Forschungszentrum OPTIMAS,
Technische Universit\"at Kaiserslautern, 67663 Kaiserslautern,
Germany}
\author{B. Hillebrands}
\affiliation{Fachbereich Physik and Forschungszentrum OPTIMAS,
Technische Universit\"at Kaiserslautern, 67663 Kaiserslautern,
Germany}

\date{\today}

\begin{abstract}
We report on the observation of reverse Doppler effect in backward spin waves reflected off of surface
acoustic waves. The spin waves are excited in a yttrium iron garnet (YIG) film. Simultaneously, acoustic
waves are also generated. The strain induced by the acoustic waves in the magnetostrictive YIG film
results in the periodic modulation of the magnetic anisotropy in the film. Thus, in effect, a travelling
Bragg grating for the spin waves is produced. The backward spin waves reflecting off of this grating
exhibit a {\it reverse} Doppler shift: shifting {\it down} rather than up in frequency when reflecting
off of an approaching acoustic wave. Similarly, the spin waves are shifted up in frequency when
reflecting from receding acoustic waves.

\end{abstract}


\maketitle%

The Doppler effect (or Doppler shift) is a well known phenomenon in which a wave emitted from a moving
source or reflected off of a moving boundary is shifted in frequency \cite{Doppler, Doppler-2}. When the
source or reflector is approaching the receiver, the frequency of received wave is shifted up in
frequency. Similarly, the frequency shifts down if the source or reflector is moving away from the
observer. The effect is widely used in radar systems, laser vibrometry and astronomical observations.

In, so called, left-handed media \cite{Pafomov} the reverse (or anomalous) Doppler effect occurs
\cite{Veselago1, Pafomov,  sceince, Reed, Leong, Stancil}. This effect is characterized by the opposite
frequency shift: waves reflect from an approaching boundary with lowered frequency. Conversely, waves
reflect from a receding boundary with higher frequency. The explanation for the reversal Doppler shift
is that in left-handed media, the group and phase velocities of the waves are in opposite directions
\cite{Veselago2}. The frequency at which the reflector produces waves is determined by the rate at which
it encounters the wave crests from the source. For a wave group approaching the reflector in a
left-handed medium, the wave crests are actually moving away from the reflector. Thus, the reflector
encounters fewer (more) crests per second if it is moving towards (away from) the source than if it were
stationary, resulting in a lower (higher) frequency of the reflected wave.

\begin{figure}[t]
\begin{center}
\scalebox{1}{\includegraphics[width=7.5 cm,clip]{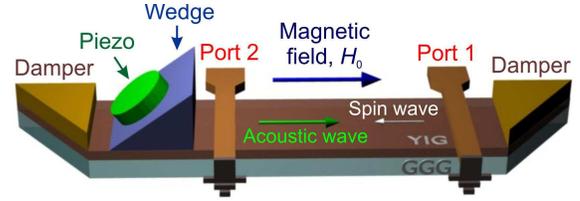}}
\end{center}
\vspace*{-0.4cm}\caption{(Color online) Experimental setup. Spin waves are excited and received in the
YIG film by stripline antennae (Port~1 and Port~2). The SAW is excited on the YIG/GGG substrate by a
piezoelectric quartz crystal and an acrylic wedge transducer.} \label{Setup}
\end{figure}

Magnetostatic spin waves travelling in a thin film magnetic material, saturated by a magnetic field
along the direction of propagation, are known to have negative dispersion. That is, the phase velocity
and group velocity are in opposite directions. Such waves are termed backward volume magnetostatic waves
(BVMSW) \cite{Damon}. Stancil {\it et al.} previously observed the reverse Doppler effect in BVMSW for
the case where the receiver is moving relative to the source \cite{Stancil}. We report here the
observation of a reverse Doppler effect for BVMSW reflecting off of a moving target, namely a travelling
surface acoustic wave. These results are interesting for both fundamental research on linear and
nonlinear wave dynamics, magnon-phonon interactions and for signal processing in the microwave frequency
range. Microwave devices such as frequency shifters, adaptive matched filters and phonon detectors may
be conceived using inelastic scattering of spin waves on acoustic waves.

The experiments were performed using 6\,$\mu$m-thick yttrium iron garnet (YIG) films, which were
epitaxially grown on 500 $\mu$m-thick, (111) oriented gadolinium gallium garnet (GGG) substrates. The
substrates were cut into strips approximately 3 mm wide and 2 cm long. To produce the conditions for
backward volume magnetostatic wave propagation, an external bias magnetic field of $H_0=1640$~Oe was
applied in the plane of the YIG film strip along its length and parallel to the direction of spin-wave
and SAW propagation (see Fig.~\ref{Setup}). BVMSWs were excited and detected in the YIG film using
microwave stripline antennae spaced 8 mm apart (shown as Port~1 and Port~2 in Fig.~\ref{Setup}). The
spin waves were generated by driving the antennae with the microwave source of a network analyzer (model
Agilent N5230C). The microwave signal power, at 1\,mW, was low enough to avoid non-linear processes. The
microwave frequency was swept through the range 6.4-6.6 GHz. Simultaneously, surface acoustic waves were
launched to propagate along the same path on the YIG/GGG sample. Longitudinal compressional waves at
frequency $f_\mathrm{SAW}=10$~MHz were generated using a piezoelectric quartz crystal and coupled to
surface modes in the YIG/GGG with an acrylic wedge transducer \cite{Hanna2}. The wedge was machined to
51$\mathrm{^o}$ for most efficiently transforming bulk acoustic waves into surface acoustic waves. A
transformer and resonant circuit were used for impedance matching between the 50 $\Omega$ source and the
piezoelectric crystal. The ends of the YIG/GGG sample were cut at a 45$\mathrm{^o}$  angle and coated
with a silicone acoustic absorber to avoid reflections (see Fig.~\ref{Setup}).

\begin{figure}[t]
\begin{center}
\scalebox{1}{\includegraphics[width=7.5 cm,clip]{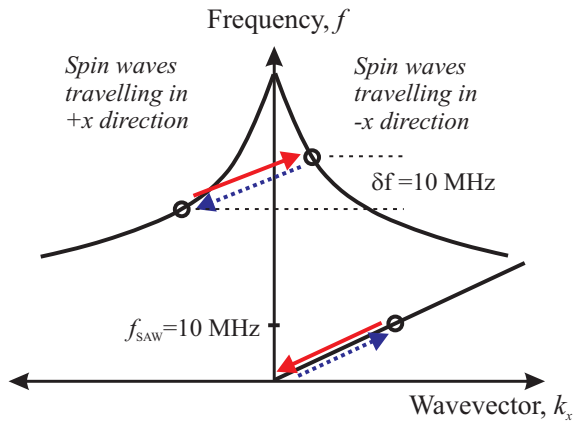}}
\end{center}
\vspace*{-0.4cm}\caption{(Color online) Schematic of dispersion curves for BVMSW and SAW. Circles
indicate the waves that participate in Bragg scattering. Red solid arrows show the process of scattering
of BVMSW on {\it co-propagating} SAW resulting in spin-waves shifted up in frequency while a phonon is
annihilated. Blue dashed arrows show the process of scattering of BVMSW on {\it counter-propagating} SAW
with the resulting spin-wave frequency shifted down while a phonon is generated.} \label{Scheme}
\end{figure}

The acoustic waves interact with the spin waves through the magnetostrictive effect in the magnetic
material \cite{Hanna1, Gulyaev}. The strain of the acoustic wave thereby periodically modulates the
magnetic properties of the film, effectively producing a travelling Bragg grating off of which the spin
waves are reflected. Fig.~\ref{Scheme} shows schematically the dispersion curves for both the BVMSW and
SAW. One can see that the group velocity of BVMSW, as determined from the slope of the dispersion curve,
is negative for positive wave vectors and vice versa. Thus, points on the BVMSW curve to the left of the
axis represent waves propagating or carrying energy to the right from Port~2 to Port~1. Conversely, spin
waves propagating to the left from Port~1 to Port~2 appear on the right side of the plot. The surface
acoustic waves have a normal, linear dispersion relation: SAW travelling to the right from the prism are
indicated by points on the right side of the plot.

The scattering process of spin waves on the acoustic waves must conserve energy and momentum.
Fig.~\ref{Scheme} shows schematically the transitions allowed by the conservation laws. The annihilation
of a phonon (red solid arrows in Fig.~\ref{Scheme}) corresponds to the generation of a magnon of higher
frequency and travelling in the opposite direction of the original spin wave. It is clear that for the
experimental setup shown in Fig.~\ref{Setup}, this interaction can be realized only for the spin wave
which propagates in the $+x$ direction, {\it i.e.}, in the same direction as the SAW. One can see that
the Doppler effect is reversed since the reflected spin wave has higher frequency. Another process is
realized with the generation of the phonon (blue dashed arrows in Fig.~\ref{Scheme}), which corresponds
to the generation of a magnon of lower frequency travelling in the opposite direction of the original
spin wave. This process takes place between counter-propagating spin and acoustic waves. The Doppler
shift, $\delta f$, is equal to the SAW frequency in both cases.

\begin{figure}[t]
\begin{center}
\scalebox{1}{\includegraphics[width=8 cm,clip]{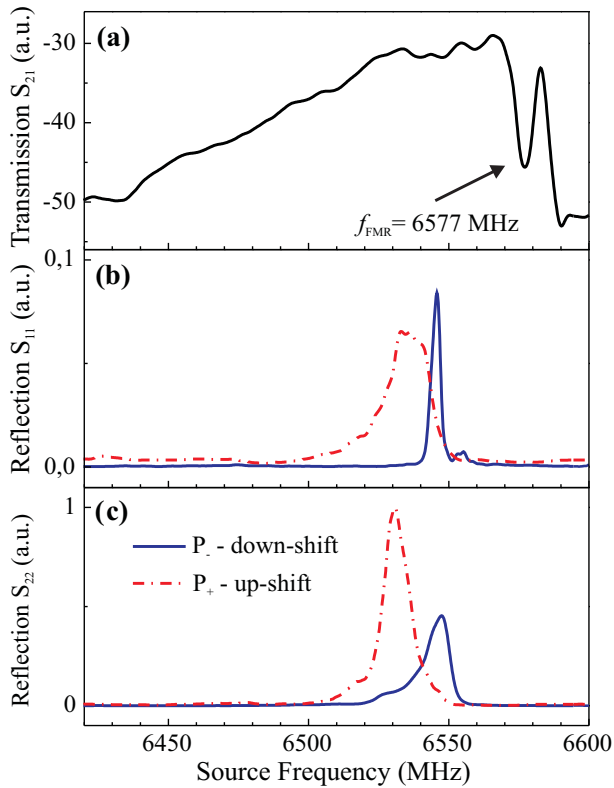}}
\end{center}
\vspace*{-0.4cm}\caption{(Color online) (a) BVMSW transmission characteristic for the YIG film. (b) and
(c) The reflection characteristics for Port~1 and Port~2 respectively. The solid blue curve, $P_{+}$, is
for the detector frequency set 10 MHz below the source frequency. The dashed red curve, $P_{-}$, is for
the detector frequency set 10 MHz above the source frequency. In each case, the frequency axis is the
swept source frequency.} \label{Result}
\end{figure}

Fig.~\ref{Result}(a) shows the experimentally measured BVMSW transmission characteristic for the YIG
film as determined from the $\mathrm{S}_{21}$ parameter (power received at Port~2 relative to the power
delivered to Port~1). The spin-wave transmission band is bounded above by the ferromagnetic resonance
frequency and below by the antenna excitation efficiency. It has a maximum just below the point of
ferromagnetic resonance ($f_\mathrm{FMR}=6577$~MHz).


Fig.~\ref{Result}(b) shows the reflection characteristics for Port~1 ($\mathrm{S}_{11}$ parameter) due
to spin waves generated at Port~1 being reflected back to the same antenna. The Doppler shifted
frequencies were measured by tuning the network analyzer to detect signals at frequencies offset by plus
and minus 10 MHz ({\it i.e.}, $\pm f_\mathrm{SAW}$) from the swept source frequency. Similarly, the
reflection characteristics for Port~2 are shown in Fig.~\ref{Result}(c). In each case, the frequency
axis is the swept source frequency.

One can see from Fig.~\ref{Result}(b) that both up and down-shifted frequencies exist for the reflected
spin waves ($P_{+}$ and $P_{-}$ signals in figure). The reason is as follows: with the microwave signal
applied to Port~1, the antenna excites spin waves propagating outwards in both directions from the
antenna. The spin waves propagating to the left, towards the acoustic source, encounter approaching
surface acoustic waves and are partially scattered back towards the source antenna with a reverse
Doppler shift down in frequency. The spin waves propagating to the right, away from the acoustic source,
encounter receding acoustic waves and are scattered back with an up-shift in frequency due to the
reverse Doppler effect.

The allowed transitions shown in Fig.~\ref{Scheme} are equivalent to the Bragg reflection conditions.
For frequencies meeting these conditions, the reflected spin wave power is maximized. Thus, the peaks in
the $P_{+}$ and $P_{-}$ curves correspond to the phonon annihilating up-shift and phonon generating
down-shift processes respectively.  The down-shift process must start at a higher spin wave source
frequency and, in the reverse transition, the up-shift process must start from a lower original spin
wave frequency. Thus, the difference in source frequency for the up-shift and down-shift process,
$\delta f$,  should be equal to the SAW frequency, $f_\mathrm{SAW}$. This is seen in the experimental
results shown in Fig.~\ref{Result}(b): the source frequency at which the $P_{+}$ signal reaches a
maximum is 10~MHz lower as compared to the $P_{-}$ signal. The peak of the down-shifted reflection is
larger and narrower because the path length over which the acoustic and spin waves can interact is
approximately two times longer on the left side of the antenna. Although both up-shifted and
down-shifted signals are present in the experimental results, it is clear from the relative amplitudes
that the up-shifted signal is due to the co-propagating waves, verifying the reverse Doppler effect.

The reflection characteristics for Port~2 are presented in Fig.~\ref{Result}(c). One can see that for
Port~2, the up-shifted reflection is stronger because the right-wards propagating spin waves (which
reflect off of receding acoustic waves) have a longer interaction path length. Note the difference in
the scale between Fig.~\ref{Result}(b) and (c); the Port~2 signals are stronger because the acoustic
amplitude is larger near the source. Similar to Port~1 results, the positions of the peaks differ by
approximately $f_\mathrm{SAW}$. The frequency difference between the maxima in $P_{-}$ and $P_{+}$
signals is $\delta f_{1}=10$~MHz and $\delta f_{2}=15$~MHz for Port~1 and Port~2 respectively. The
discrepancy in $\delta f_{2}$ from the expected 10~MHz is likely due to part of the interaction
occurring under the wedge: here, the SAW wavelength differs from that in the unloaded YIG film. Thus,
the spin wave frequencies at which the Bragg conditions are met are different for the co- and
counter-propagating cases. It should be noted that the actual Doppler frequency shift in the reflected
spin wave is exactly 10 MHz in both cases, as determined by the detector frequency offset.

To construct a simple and representative theoretical model, we consider the BVMSW dispersion relation to
be nearly linear for small wavenumbers ($kd \ll 1$, where $d$ is the thickness of the YIG film). Thus,
we can write
\begin{equation}\label{f_SW}
f_\mathrm{SW} (k) = f_\mathrm{FMR} + \upsilon_\mathrm{SW} \cdot k, \nonumber
\end{equation}
where
\begin{equation}\label{v_SW}
\upsilon_\mathrm{SW} = - \frac{f_\mathrm{H} f_\mathrm{M}}{4 f_\mathrm{FMR}} \cdot d \nonumber
\end{equation}
is the group velocity of BVMSW. Here $f_\mathrm{H} = \gamma H_0$, $f_\mathrm{M} = 4 \pi \gamma M_0$,
where $\gamma = 2.8$~MHz/Oe is the gyromagnetic ratio.

The dispersion relation for the SAW is nearly linear with $f_\mathrm{SAW} (k) = \upsilon_\mathrm{SAW}
\cdot k$, where $\upsilon_\mathrm{SAW}$ is the phase and group velocity of the acoustic wave. Fulfilling
laws of energy and momentum conversation by the transitions indicated in Fig.~\ref{Scheme}, a simple
equation can be derived for the initial spin-wave frequencies $f_{+}$ and $f_{-}$ which corespond to the
maxima of $P_{+}$ and $P_{-}$:

\begin{equation}\label{res}
f_{\pm} = f_\mathrm{FMR} - \frac{f_\mathrm{SAW}}{2}(\frac{\upsilon_\mathrm{SW}}{\upsilon_\mathrm{SAW}}
\pm 1) \nonumber \end{equation}

Using saturation magnetization $4\pi M_0 = 1750$~G for the YIG film, BVMSW group velocity
$\upsilon_\mathrm{SW} = 3.2$~cm/$\mu$s, and SAW velocity $\upsilon_\mathrm{SAW} = 0.5$~cm/$\mu$s this
equation gives the values for $f_{+} = 6537$~MHz and $f_{-} = 6547$~MHz which is in good agreement with
the experimental data (see Fig.~\ref{Result}).

In conclusion, we have observed the reverse Doppler effect in backward spin waves reflected off of
surface acoustic waves. Both possible situations were analyzed: the scattering of BVMSW from
co-propagating and counter-propagating SAW. It was shown that the frequencies of scattered spin waves in
both cases were shifted by the frequency of SAW according to the reverse Doppler effect. The results are
in good agreement with the theoretical analysis based on the dispersion curves of spin waves and
acoustic waves. Similar reverse Doppler effects are to be expected in other left-handed media.

This work was partially supported by the DFG SE 1771/1-1, and NSF ECCS 0645236. Special acknowledgments
to Prof. G. A. Melkov for valuable discussions.

\end{document}